\documentclass{conm-p-l}

\copyrightinfo{2013}{}

\usepackage{graphicx}
\usepackage{subfigure}

\usepackage{amssymb,amsmath,amsthm,amscd}

\theoremstyle{definition}

\theoremstyle{remark}

\numberwithin{equation}{section}

\newcommand{\e}{\epsilon}

\renewcommand{\k}{\kappa}
\newcommand{\ga}{\gamma}

\newcommand{\Dl}{\Delta}

\newcommand{\ra}{\rightarrow}
\newcommand{\al}{\alpha}
\newcommand{\be}{\beta}
\newcommand{\sg}{\sigma}

\newcommand{\pa}{\partial}

\newcommand{\na}{\nabla}

\begin{document}

\title[Major open problems in chaos theory and nonlinear dynamics]{Major open problems in chaos theory and nonlinear dynamics}

\author{Y. Charles Li}
\address{Department of Mathematics, University of Missouri, 
Columbia, MO 65211, USA}
\email{liyan@missouri.edu}
\urladdr{http://www.math.missouri.edu/~cli}

\thanks{}

\subjclass{Primary 37, 76, 92, 70; Secondary 34, 35, 82, 80}
\date{}

\dedicatory{}


\begin{abstract}
Nowadays, chaos theory and nonlinear dynamics lack research focuses. Here we mention a few major 
open problems: 1. an effective description of chaos and turbulence, 2. rough dependence on initial data,
3. arrow of time, 4. the paradox of enrichment, 5. the paradox of pesticides, 6. the paradox of plankton.
\end{abstract}

\maketitle

\section{Introduction}

Chaos theory originated from studies in classical mechanics (H. Poincar\'e \cite{Poi92}), fluid mechanics 
(E. Lorenz \cite{Lor63}), and ecology (R. May \cite{May76}). By now chaos theory has spread to almost 
every scientific area and beyond. Overall, chaos is understood but not tamed. In fact, it is not clear 
whether or not it is tractable! More specifically, the mechanism of how chaotic dynamics operates is 
understood; how to effectively describe chaos in term of some sort of averaging is beyond reach (i.e. 
not tamed); it is not even clear what kind of averaging mean we should be  after for! These questions will 
form the first major open problem to be discussed below. Returning to the three specific areas where 
chaos theory originated, chaotic dynamics in classical mechanics is understood; chaotic dynamics 
(turbulence) in fluid mechanics is being understood in infinite dimensional phase space under the 
flow defined by Navier-Stokes equations; chaotic dynamics (and nonlinear dynamics in general) in ecology 
is not (or poorly) understood. In classical mechanics, dynamics is generally governed by a 
system of finitely many ordinary differential equations, and numerical simulations of such a system 
have very good precision and can in principle reveal all the detailed structures in the finite dimensional 
phase space. In fluid mechanics, dynamics is governed by Navier-Stokes equations. The phase space
is infinite dimensional. Numerical simulations on Navier-Stokes equations have a long way from 
accuracy. Nevertheless, explorations on nonwandering structures such as fixed points (steady states), 
periodic orbits, homoclinic orbits, and bifurcations to chaos have established reliable results 
\cite{Nag90} \cite{Vis07} \cite{VK11} \cite{KE12}. Philosophically speaking, chaos in fluid is a part of 
turbulence. The question is whether or not turbulence contains more. This question will be  addressed in 
the second open problem below: rough dependence on initial data. The initiation (onset) of turbulence is 
much trickier \cite{LL11} \cite{LL13} than that of chaos in finite dimensions. 
In ecology \cite{May99}, the accuracy of various mathematical models is difficult to test. Large 
scale field studies are difficult to conduct. These difficulties lead to the poor understanding of 
nonlinear dynamics in ecology.

Chaos theory forms the core of a greater area called complex systems. Even though there has 
been tremendous effort in the grand area of complex systems, no substantial new scientific result has been
obtained. Nowadays, research in the area of chaos theory lacks focus, partly due to its spreading 
into the area of complex systems, and partly due to its spreading into almost all scientific areas. Here 
we shall mention a few major open problems which we believe to be of fundamental importance in 
chaos theory and nonlinear dynamics. 

\section{An effective description of chaos and turbulence}

The search for an effective description of turbulence started from Reynolds average \cite{Rey95} 
\cite{Hin75}. Reynolds average was designed for a stochastic signal which is the sum of an average 
signal and a small stochastic perturbative signal. Unfortunately chaos and turbulence are far from such 
stochastic signals. Thus Reynolds average is far from an effective description of turbulence. Applying 
Reynolds average to chaos and turbulence will inevitably lead to an unsolvable closure problem. 
Since turbulence is chaos under Navier-Stokes flow, it is a good idea to first search for an effective 
description for chaos of simple systems. In fact, one can view fractal dimension \cite{Dev03} 
and SRB measure \cite{ASY97} for strange attractor as description of chaos. Fractal dimension does 
not have much engineering value. SRB measure is elusive for most of systems. Thus neither 
fractal dimension nor SRB measure is an effective description of chaos. Horseshoe \cite{Sma67} 
\cite{Li99} and shadowing \cite{Pal88} \cite{Li03} in combination with Bernoulli shift are also 
descriptions of chaos. Shadowing has the greatest potential to generate an effective description 
of chaos \cite{Li07} \cite{LL10}. But with the current computer capacity, shadowing of the entire 
turbulent attractor is still beyond reach. A description of turbulence which is effective in engineering 
means a solution of the problem of turbulence. 

\section{Rough dependence on initial data}

The signature of chaos is "sensitive dependence on initial data", here I want to address "rough dependence on initial data"
which is very different from sensitive dependence on initial data. For solutions (of some system) that exhibit sensitive dependence on initial data,
their initial small deviations usually amplify exponentially (with an exponent named Liapunov exponent), and it takes time for the deviations
to accumulate to substantial amount (say order $O(1)$ relative to the small initial deviation). If $\e$ is the initial small deviation, and $\sg$ is the
Liapunov exponent, then the time for the deviation to reach $1$ is about $\frac{1}{\sg} \ln \frac{1}{\e}$. On the other hand, for solutions  that exhibit rough dependence on initial data, their initial small deviations can reach substantial amount instantly. Take the 3D or 2D Euler equations of fluids as
the example, for any $t \neq 0$ (and small for local existence), the solution map that maps the initial condition to the solution value at time $t$ is
nowhere locally uniformly continuous and nowhere differentiable \cite{Inc13}. In such a case, any small deviation of the initial condition can
potentially reach substantial amount instantly. My conjecture is that the high Reynolds number violent turbulence is due to such
rough dependence on initial data, rather than sensitive dependence on initial data of chaos. When the Reynolds number is sufficiently large (
the viscosity is sufficiently small), even though the solution map of the Navier-Stokes equations is still differentiable, but the derivative of the
solution map should be extremely large everywhere since the solution map of the Navier-Stokes equations becomes the solution map of the
Euler equations when the viscosity is zero (the Reynolds number is set to be infinity). Such everywhere large derivative of the solution
map of the Navier-Stokes equations should manifest itself as the development of violent turbulence in a short time. In summary, not large
enough Reynolds number turbulence may be due to sensitive dependence on initial data of chaos, while large enough Reynolds number
turbulence may be due to rough dependence on initial data.

The type of rough dependence on initial data shared by the solution map of the Euler equations is difficult to find in finite dimensional systems.
The solution map of the Euler equations is still continuous in initial data. Such a solution map (continuous, but nowhere locally uniformly continuous)
does not exist in finite dimensions. This may be the reason that one usually finds chaos (sensitive dependence on initial data) rather than
rough dependence on initial data. If the  solution map of some special finite dimensional system is nowhere continuous, then the dependence
on initial data is rough, but may be too rough to have any realistic application.

\section{The arrow of time}

\begin{figure}[h] 
\centering
\subfigure[$t=2.59$]{\includegraphics[width=2.3in,height=1.15in]{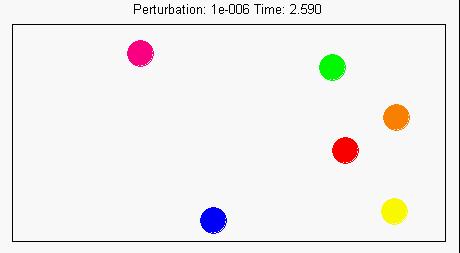}}
\subfigure[$t=3.286$]{\includegraphics[width=2.3in,height=1.15in]{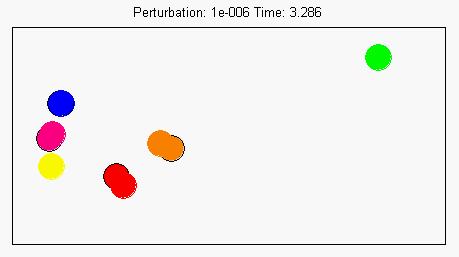}}
\subfigure[$t=3.42$]{\includegraphics[width=2.3in,height=1.15in]{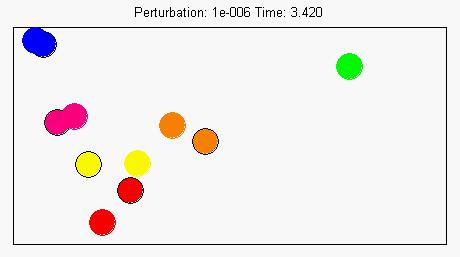}}
\subfigure[$t=3.954$]{\includegraphics[width=2.3in,height=1.15in]{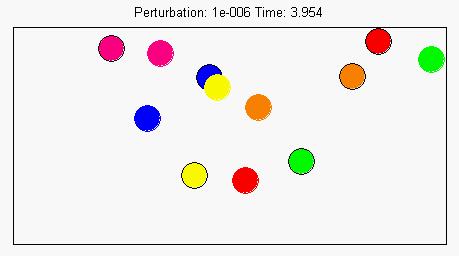}}
\caption{The evolution of the six disks and the evolution of the perturbed six disks. Since the perturbation 
size is $10^{-6}$, initially the unperturbed and the perturbed disks coincide almost completely. The radius of the disk 
is $0.25$, and the rectangle domain is $8\times 4$.}
\label{pan1}
\end{figure}

\begin{figure}[h] 
\centering
\subfigure[$t=3.608$]{\includegraphics[width=2.3in,height=1.15in]{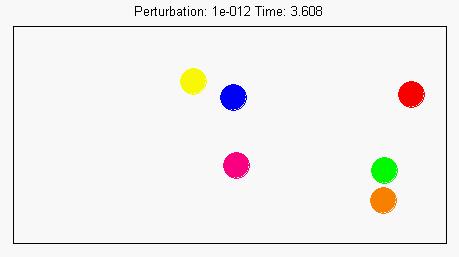}}
\subfigure[$t=4.046$]{\includegraphics[width=2.3in,height=1.15in]{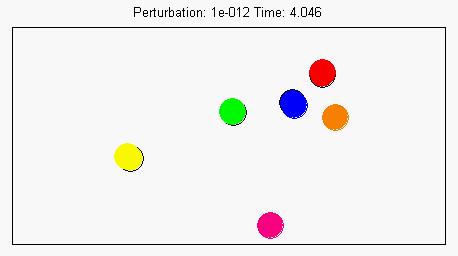}}
\subfigure[$t=4.184$]{\includegraphics[width=2.3in,height=1.15in]{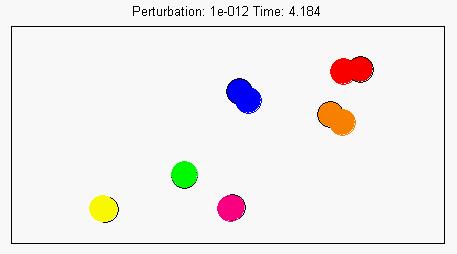}}
\subfigure[$t=5.248$]{\includegraphics[width=2.3in,height=1.15in]{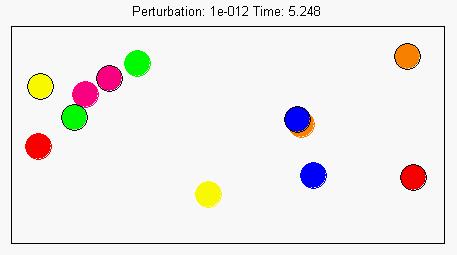}}
\caption{The same setup as in Figure \ref{pan1} except that the perturbation 
size is $10^{-12}$.}
\label{pan2}
\end{figure}

\begin{figure}[ht] 
\centering
\includegraphics[width=4.5in,height=1.5in]{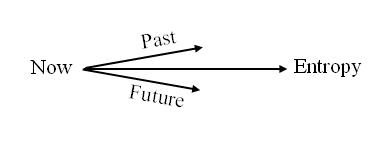}
\caption{The diagram of the arrow of time. The `Past', `Now' and `Future' are coordinate time, and 
the `Entropy' is the thermodynamic equilibrium entropy.}
\label{DA}
\end{figure}

The arrow of time generally refers to the phenomenon of microscopic time reversibility v.s. 
macroscopic time irreversibility. That is, macroscopically there is an arrow of time. The arrow of time
is an important problem in several branches of physics. We believe that it is also an important 
problem in chaos theory. A simple example of the arrow of time in chaos theory is the problem of 
releasing bouncing balls in the half box to the whole box. Due to the chaotic dynamics of the bouncing 
balls, after sufficiently long time, even though every ball's velocity is simultaneously reversed, the 
chance of all the balls simultaneously return to the half box is very small ($2^{-N}$ where $N$ is the 
number of balls) \cite{LY12}.  The inevitable perturbations amplify substantially via chaotic dynamics 
after enough time, see Figures \ref{pan1} and \ref{pan2}. The chaotic dynamics liberates the mathematical 
control of the Newtonian law 
to the balls so that after sufficiently long time, the orbits of the balls loose the memory of the initial 
condition, and are far away from the purely mathematical orbits! When the number of balls increases, 
e.g. the case of gas molecules, the problem becomes a problem of thermodynamics. In thermodynamics,
the arrow of time refers to the second law of thermodynamics in which the entropy can only change in 
one direction (i.e. the time's arrow). Our diagram of the arrow of time in thermodynamics is shown in 
Figure \ref{DA}. The term, arrow of time, was introduced by Arthur Eddington in 
1927. Now several types of arrows of time have been studied. These include thermodynamic, 
cosmological, psychological, and causal arrows of time. Cosmological arrow of time means the universe's 
expansion, psychological arrow of time means that one can only remember the past not the future, and 
causal arrow of time means that cause precedes its effect. The mechanisms of these different arrows of time 
may be different. Chaos theory seems to be most directly relevant to the thermodynamic arrow of time. For 
the cosmological arrow of time, one may ask the question: Is the universe still expanding if the velocity of 
every object (or every molecule) is simultaneously reversed? If the answer is yes, then chaos theory may still 
be relevant. Even though brain dynamics may be chaotic, the direct relevance of chaos theory to the 
psychological arrow of time is not clear. The relevance of chaos theory to the causal arrow of time is even 
more unclear. 

\section{The paradox of enrichment}

The paradox of enrichment was first observed by M. Rosenzweig \cite{Ros71} in a class of mathematical models 
on the dynamics of predators and prey. The paradox roughly says that the class of mathematical models predicts 
that increasing the nutrition to the prey may lead to the extinction of both the prey and the predator. The most 
important question is whether or not this paradox can be observed experimentally. It is possible that the paradox is 
purely an artifact of the mathematical models, while in reality increasing the nutrition to the prey never leads to an
extinction. If that is the case, then developing better mathematical models is necessary. 

\begin{figure}[ht] 
\centering
\includegraphics[width=4.5in,height=4.5in]{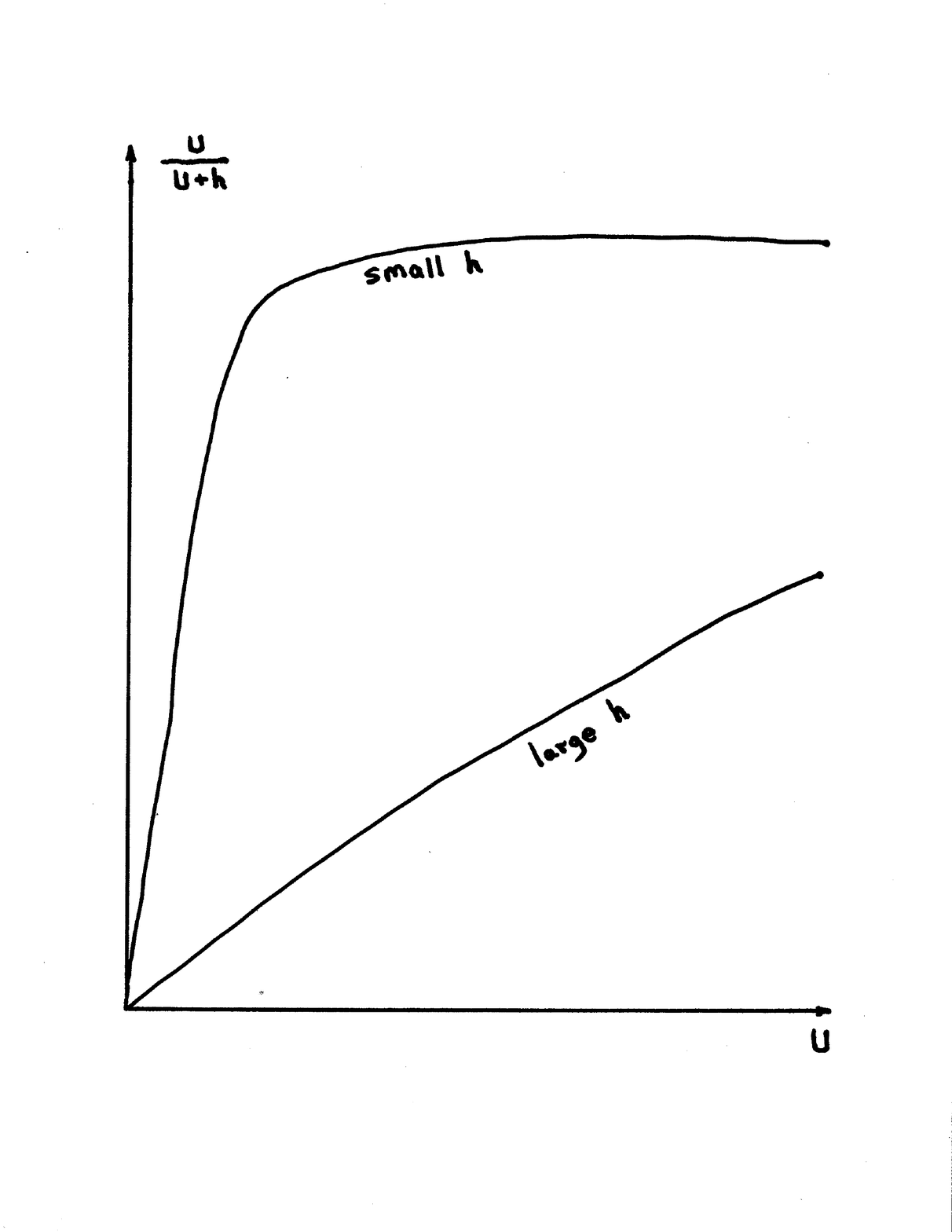}
\caption{The predation graphs.}
\label{slh}
\end{figure}

Specifically let us look at one of such mathematical models \cite{Ros71},
\begin{eqnarray}
& & \frac{d U}{d T}  = \al U \left (1- \frac{U}{b} \right ) - \ga \frac{U}{U+h} V , \label{ud1} \\
& & \frac{d V}{d T}  = \left (\k \ga \frac{U}{U+h} - \mu \right ) V , \label{ud2} 
\end{eqnarray}
where $U$ is the prey density, $V$ is the predator density, $T$ is the time coordinate, $\al$ is the maximal per 
capita birth rate of 
the prey, $b$ is the carrying capacity of the prey from the nutrients, $h$ is the half-saturation prey density for 
predation, $\ga$ is the coefficient of the intensity of predation, $\k$ is the coefficient of food utilization of the 
predator, and $\mu$ is the mortality rate of the predator. The paradox focuses upon the 
steady state given by
\[
\k \ga \frac{U}{U+h} - \mu = 0, \ \al \left (1- \frac{U}{b} \right ) - \ga \frac{1}{U+h} V = 0 .
\]
It turns out that when other parameters are fixed, increasing $b$ leads to the loss of stability of this steady state,
in which case, a limit cycle attractor around the steady state is generated. As $b$ increases, the limit cycle gets 
closer and closer to the $V$-axis. That is, along the limit cycle attractor, the prey population $U$ decreases to
a very small value. Under the ecological random perturbations, $U$ can reach $0$, i.e. extinction of the prey. With 
the extinction of the prey, the predator will become extinct soon. On the other hand, increasing $b$ means increasing 
the carrying capacity of the prey, which can be implemented by increasing the prey's nutrients, i.e. enrichment of 
the prey's environment. Intuitively, increasing $b$ should enlarge the prey population and make it more robust from 
extinction. This is the paradox of enrichment. In order to resolve the paradox of enrichment, it is fundamental to 
rewrite the system (\ref{ud1})-(\ref{ud2}) in the dimensionless form \cite{FL13}:
\begin{eqnarray}
& & \frac{d u}{d t} = u(1-u ) - \frac{u}{u+H} v , \label{UMd1} \\
& & \frac{d v}{d t} = k\left (\frac{u}{u+H} - r \right ) v , \label{UMd2} 
\end{eqnarray}
where $u = U/b$, $v= V \ga /(\al b)$, $t=\al T$, and the dimensionless numbers are given by
\begin{equation}
H = \frac{h}{b}, \ r = \frac{\mu}{\k \ga}, \ k = \frac{\k \ga}{\al}.
\label{dln}
\end{equation}
We name $H$: the capacity-predation number, and $r$: the mortality-food number. 
\begin{itemize}
\item {\bf The Resolution} \cite{FL13}: Unlike the original form of the model (\ref{ud1})-(\ref{ud2}), the dimensionless form 
of the model (\ref{UMd1})-(\ref{UMd2}) is governed by $3$ dimensionless numbers $H$, $r$ and $k$ (\ref{dln}). $H$ is a ratio 
of the half-saturation 
$h$ and carrying capacity $b$, while $r$ and $k$ are independent of $h$ and $b$. Increasing the carrying capacity $b$ (for fixed half-saturation $h$) and 
decreasing the half-saturation $h$ (for fixed carrying capacity $b$) have the same effect on the capacity-predation number 
$H$, that is, $H$ decreases. Decreasing the half-saturation $h$ implies more aggressive predation (especially when the 
prey population $U$ is small), see Figure \ref{slh}. Notice that 
\[
\text{As } U \ra 0^+, \ \frac{U}{U+h} \ra 1;
\]
and 
\[
\frac{d}{dU} \frac{U}{U+h} \bigg |_{U=0} = 1/h .
\]
Since there is no paradox between more aggressive predation (especially when the prey population $U$ is small) and 
extinction of prey, the paradox of enrichment now reduces to a 
paradox between more aggressive predation (decreasing the half-saturation $h$) and enrichment (increasing the carrying 
capacity $b$). As mentioned above, the special feature of the model (\ref{UMd1})-(\ref{UMd2}) is that more aggressive 
predation (decreasing $h$) and enrichment (increasing $b$) is not a paradox, and results in the same effect on the governing 
dimensionless number $H$. This offers a resolution to the so-called paradox of enrichment. 
\end{itemize}
From the above analysis, we can see that the paradox is purely generated by the particular mathematical model. If we
replace the predation term $\frac{U}{U+h}$ in both (\ref{ud1}) and (\ref{ud2}) by $U$ (a mild predation when $U$ is small),
the paradox disappears.

\section{The paradox of pesticides}

Unlike the paradox of enrichment, the paradox of pesticides was observed in experiments \cite{HMCVJ07} \cite{Ham08} \cite{LTH98}.
The paradox of pesticides says that pesticides may dramatically increase the population of a pest when 
the pest has a natural predator. Right after the application of the pesticide, of course the pest 
population shall decrease (so shall the predator). But the pest may resurge later on in much more 
abundance resulting in a population well beyond the crop's economic threshold. Roughly speaking, the pesticide 
reduces the populations of both the pest and its predator, and the ratio of the population of the pest to the population of the
predator is changed so that the resurgent pest population can be much more in abundance. To guide the application of 
pesticide in such a circumstance, a good mathematical model will be important. From the perspective of mathematical models,
the phenomenon can be easily understood \cite{LY13}. To model the effect of pesticides on pest resurgence, a simple 
mathematical model is the Lotka-Volterra system with forcing,
\begin{eqnarray}
\frac{dH}{dt} &=& H(a-bP) - \al \Dl (t-T), \label{LVF1} \\
\frac{dP}{dt} &=& P(bcH-d) - \be \Dl (t-T), \label{LVF2}
\end{eqnarray}
where $H$ is the pest population, $P$ is the pest's predator population, ($a$, $b$, $c$, $d$, $\al$, $\be$) are 
positive constants, $\Dl (t)$ is an approximation of the delta function, and the $\Dl (t-T)$ terms represent 
the effects of pesticides. Specifically, we choose $\Dl (t-T)$ to be
\[
\Dl (t-T) = 1/\e , \text{ when } t \in [T, T+\e]; \ \ = 0, \text{ otherwise}.
\]
\begin{figure}[ht] 
\centering
\includegraphics[width=5.5in,height=5.5in]{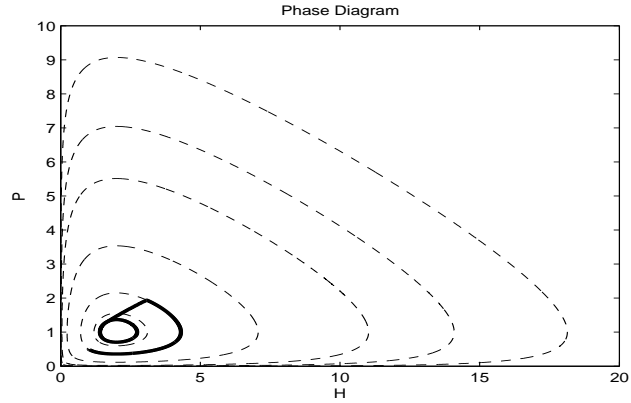}
\caption{A pesticide forced orbit of (\ref{LVF1})-(\ref{LVF2}), $a=b=d=1$, $c=0.5$, $T=3.5$, $\e = 0.07$, 
$\al = 1.4$, $\be = 0.7$. In this case, the pest population deceases and maintains at a lower amplitude oscillation.}
\label{FFD}
\end{figure}
\begin{figure}[ht] 
\centering
\includegraphics[width=5.5in,height=5.5in]{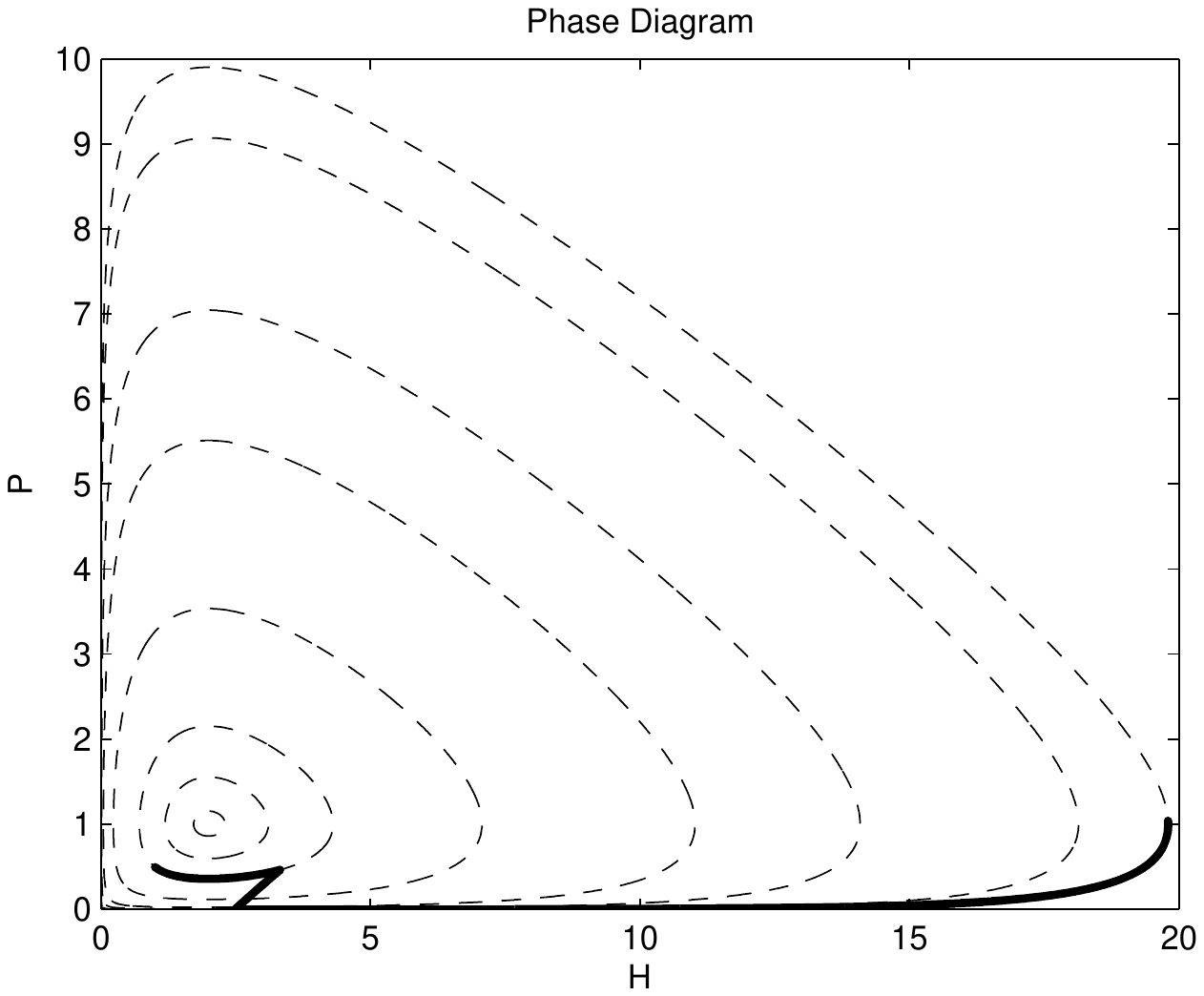}
\caption{A pesticide forced orbit of (\ref{LVF1})-(\ref{LVF2}), $a=b=d=1$, $c=0.5$, $T=2$, $\e = 0.047$, 
$\al = 0.94$, $\be = 0.47$. In this case, the pest population resurges with dramatic increase beyond the 
crop's economic threshold.}
\label{FFI1}
\end{figure}
\begin{figure}[ht] 
\centering
\includegraphics[width=5.5in,height=5.5in]{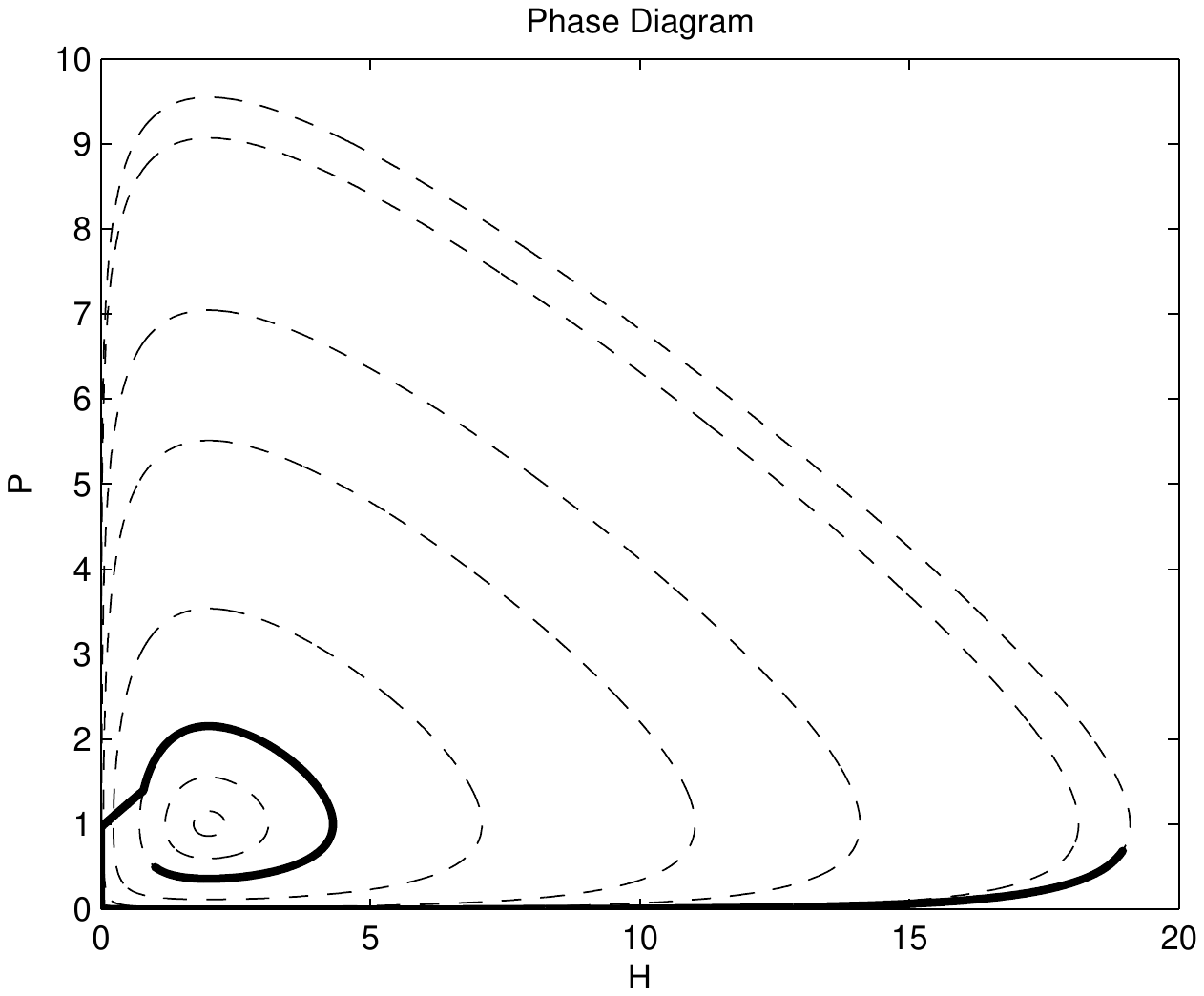}
\caption{A pesticide forced orbit of (\ref{LVF1})-(\ref{LVF2}), $a=b=d=1$, $c=0.5$, $T=5$, $\e = 0.039$, 
$\al = 0.78$, $\be = 0.39$. In this case, the pest population resurges with dramatic increase beyond the 
crop's economic threshold.}
\label{FFI2}
\end{figure}
\begin{figure}[ht] 
\centering
\includegraphics[width=5.5in,height=5.5in]{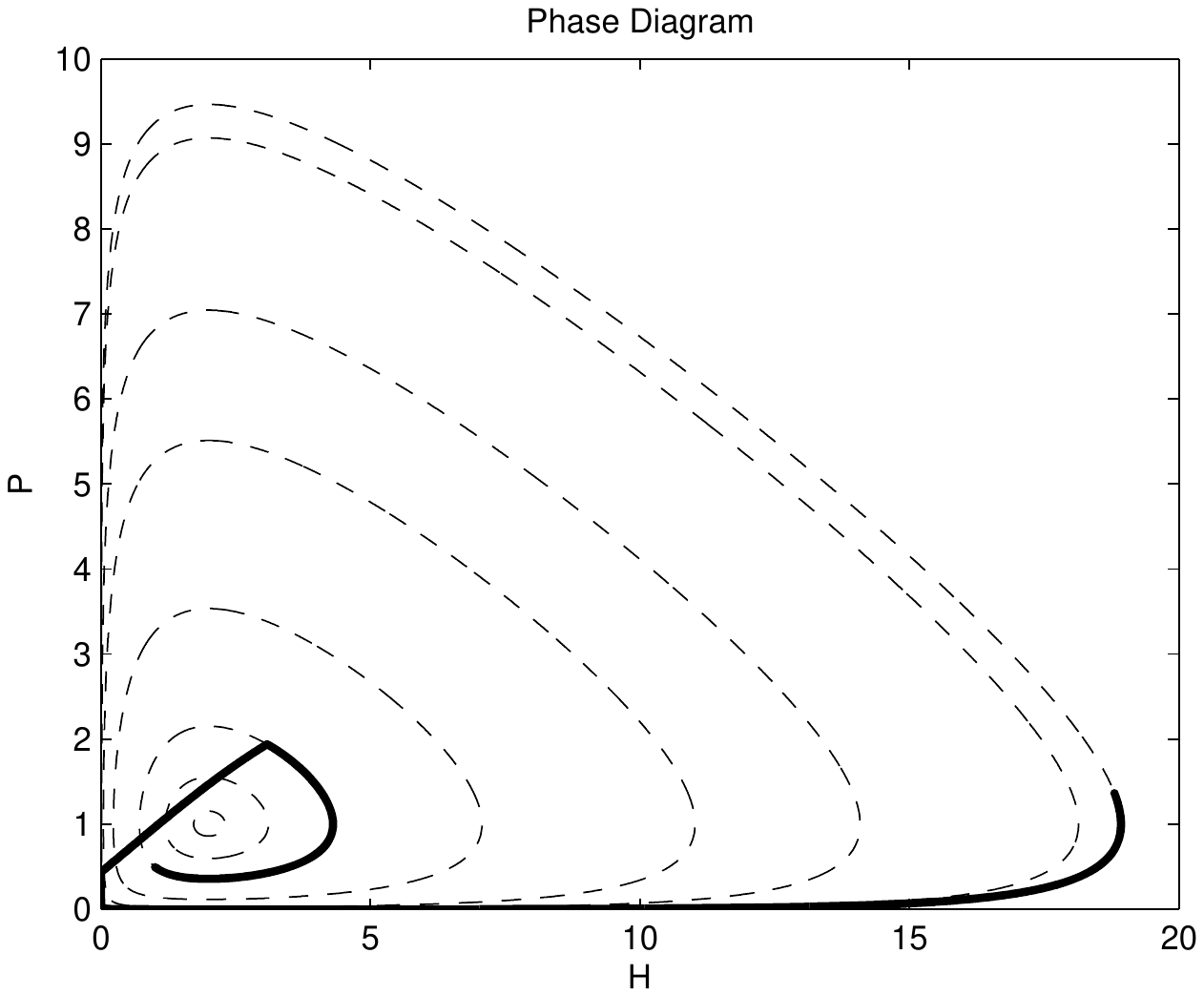}
\caption{A pesticide forced orbit of (\ref{LVF1})-(\ref{LVF2}), $a=b=d=1$, $c=0.5$, $T=3.5$, $\e = 0.149$, 
$\al = 2.98$, $\be = 1.49$. In contrast to Figure \ref{FFD}, here large amount of pesticides is applied,
and results in that the pest population resurges with dramatic increase beyond the 
crop's economic threshold.}
\label{FFI3}
\end{figure}
The key points on understanding the paradox of pesticides via the simple mathematical model are as follows:
\begin{enumerate}
\item The timing of applying the pesticides is crucial. If the pesticides are applied when the populations of both the 
pests and the predators are relatively large, then a decrease in both populations can be achieved, see Figure 
\ref{FFD}. On the other hand, if the pesticides are applied when either the pest's population or the 
predator's population is relatively small, then a dramatic increase in the resurgent pest's population occurs, leading 
to pest's population well beyond the crop's economic threshold, see Figures \ref{FFI1} and \ref{FFI2}. From these 
figures, it is clear that even though the pesticides only kill the pests rather than their predators (that is, 
right after the application of the pesticides, the pest's population decreases, while the predator's population 
maintains the same), the pests still resurge in abundance beyond the crop's economic threshold. This is because
that when the population of the pests decreases, the predator's population will decreases too, since the 
predators feed on the pests. It is the relatively minimal values of both the pest's population and the 
predator's population that decide how large the cycle which they are going to sit on in the phase plane.
\item The amount of pesticides applied is also important. In the case of Figure \ref{FFD}, if the amount 
of pesticides is large enough, a dramatic increase in the resurgent pest's population can still occur as 
shown in Figure \ref{FFI3}.
\end{enumerate}
The above conclusions also applies when spatial dependence is taken into account \cite{LY13}.

\section{The paradox of plankton}

In ecology, the Liebig's law says that population growth is controlled not by the total amount of resources available,
but by the amount of the scarcest resource (limiting factor). For instance, according to the Bateman principle, females 
spend more energy on generating offsprings than males do, thus females are a limiting resource over which males compete 
in most species. In the case of the plankton, especially phyto-plankton (in contrast to zoo-plankton), many (hundreds) 
species are competing for one or a few limiting resources (nutrients with severe deficiency in the summer). According to 
the principle of competitive exclusion,
the final equilibrium state should be taken over by one or a few species according to the limiting factors. On the other hand,
in reality hundreds species of the plankton coexist. This paradoxical question was first raised by Hutchinson \cite{Hut61}. 
Hutchinson proposed the idea that this is due to the fact that equilibrium state cannot be reached in reality. Since 
Hutchinson's work, there have been many studies on the paradox \cite{RC07}. The problem involves two branches of chaotic 
dynamics: fluids and ecology. 

Plankton drifting in (turbulent) water is an interesting problem for chaotician to model. 
Let $v(t,x)$ be the three-dimensional (turbulent) water velocity, $w_n(t,x)$ ($n=1,2, \cdots, N$) be the plankton 
drifting velocities relative to water, $x \in \mathbb{R}^3$. In reality $v(t,x)$ is governed by the 
Navier-Stokes equations. Numerical simulations of the full Navier-Stokes equations are still very challenging.
We can conveniently model the water velocity by a chaotic time evolution of spatial patterns (with vortex structure).
We believe that the essential mechanism of the plankton drifting can be captured by such a modeling.
Let $f(t,x)$ be the density of the limiting resource (deficient neutrient), and $\rho_n (t,x)$ be the densities of 
different species of plankton, 
$n=1,2, \cdots, N$. The model is given by
\begin{eqnarray}
&&\pa_t f + (v \cdot \na ) f = - G(\rho_1, \rho_2, \cdots , \rho_N)f + F(t,x), \label{feq}  \\
&&\pa_t \rho_n + [(v + w_n) \cdot \na ] \rho_n = H_n(f) \rho_n, \ \ n=1,2, \cdots , N , \label{peq}
\end{eqnarray}
where
\[
G(0,0, \cdots , 0) = 0, 
\]
and $G(\rho_1, \rho_2, \cdots , \rho_N)$ is strictly monotonically increasing in each $\rho_n$; 
\[
H_n (f^n) = 0, \ \ \text{where} \ f^n \ \text{are values of} \ f , 
\]
$H_n (f)$ is strictly monotonically increasing in $f$;
\[
w_n = \k_n \na f + \  \text{random noise  or random walk},
\]
and $\k_n$ are the drifting coefficients of different plankton. 

Through $G(\rho_1, \rho_2, \cdots , \rho_N)$, different food consumption rates by different species of 
plankton can be introduced. $F(t,x)>0$ represents food regeneration, $F(t,x)$ is periodic in $t$ with 
relatively long period. $f^n$ are the `critical' densities of food for different species of plankton. 
$H_n(f)$ model the growth rates from nutrition of different species of plankton.

\end{document}